\documentclass[aps,pra,twocolumn,floatfix,showpacs,preprintnumbers,amsmath,amssymb,10pt]{revtex4-1}
\usepackage[dvips]{graphics}

\DeclareMathAlphabet{\mathsfit}{\encodingdefault}{\sfdefault}{m}{sl}
\SetMathAlphabet{\mathsfit}{bold}{\encodingdefault}{\sfdefault}{bx}{sl}

\usepackage{mathptmx}
\usepackage{psfrag}
\usepackage{graphicx}
\usepackage{bm}
\usepackage{amsmath}
\usepackage{trfsigns}
\usepackage{amssymb}
\usepackage{subfigure}
\usepackage[usenames,dvipsnames]{color}
\usepackage{dashrule}
\usepackage{textgreek}
\usepackage[english]{babel}
\usepackage{contour}
\usepackage{upgreek,bm}
\usepackage{color}
\usepackage{natbib}
\definecolor{dred}{rgb}{.6,.0,0.}
\definecolor{dblue}{rgb}{.0,.0,0.6}

\renewcommand{\vec}[1]{\mathbf{#1}}

\newcommand{\tens}[1]{\mbox{\textsf{\textbf{#1}}}}

\newcommand{\Greektens}[1]{\contour[3]{black}{#1}}

\newcommand{\dif}{\mathrm{d}}
\newcommand{\mi}{\textrm{i}} 
\newcommand{\me}{\mathrm{e}}

\begin{document}

\title{Inducing and controlling rotation on small objects using photonic topological materials}

\author{Frieder Lindel$^1$}
\author{George W. Hanson$^2$}
\author{Mauro Antezza$^{3,4}$}
\author{Stefan Yoshi Buhmann$^{1,5}$}
\affiliation{$^1$ Physikalisches Institut, Albert-Ludwigs-Universit\"at Freiburg, Hermann-Herder-Stra{\ss}e 3, 79104 Freiburg, Germany\\
$^2$ Department of Electrical Engineering, University of Wisconsin-Milwaukee, 3200 N. Cramer St., Milwaukee, Wisconsin 53211, USA\\
$^3$ Laboratoire Charles Coulomb, UMR 5221 Universit\'{e} de Montpellier and CNRS, F-34095 Montpellier, France\\
$^4$ Institut Universitaire de France, 1 rue Descartes, F-75231 Paris Cedex 05, France\\ 
$^5$ Freiburg Institute for Advanced Studies, Albert-Ludwigs-Universit\"at Freiburg, Albertstra{\ss}e 19, 79104 Freiburg, Germany}

\date{\today}

\begin{abstract} 

Photonic topological insulator plates violate Lorentz reciprocity which leads to a directionality of surface-guided modes. This in-plane directionality can be imprinted via an applied magnetic field. On the basis of macroscopic quantum electrodynamics in nonreciprocal media, we show that two photonic topological insulator surfaces are subject to a tuneable, magnetic-field dependent Casimir torque. Due to the directionality, this torque exhibits a unique $2\pi$ periodicity, in contradistinction to the Casimir torques encountered for reciprocal uniaxial birefringent media or corrugated surfaces which are $\pi$-periodic. Remarkably, the torque direction and strength can be externally driven in situ by simply applying a magnetic field on the system, and we show that this can be exploited to induce a control the rotation of small objects. Our predictions can be relevant for nano-opto-mechanical experiments and devices.

\end{abstract}

\maketitle


The Casimir force was originally proposed as an attractive force between two perfectly conducting plates due to a reduced virtual photon pressure in the space between the plates \cite{casimir_attraction_1948}. In macroscopic quantum electrodynamics (QED) the Casimir force was further generalized to bodies of arbitrary shape and material by realizing that its existence stems from fluctuating charge carriers within the material \cite{dzyaloshinskii_general_1961}. Subsequently the Casimir force for objects consisting of anisotropic materials or possessing anisotropic surfaces like birefrigent plates \cite{van_enk_casimir_1995}, magnetodielectric metamaterials \cite{rosa_casimir_2008} or corrugated metals \cite{rodrigues_vacuum-induced_2006} was studied. Since all those materials have a distinguishable axis in the plane of the plates it is natural to ask whether the Casimir energy depends on the relative angle between the two axes when bringing two anisotropic surfaces together. It turns out that indeed one obtains a torque when the angle between the axes is different from zero or $\pi$ \cite{barash_moment_1978,philbin_alternative_2008,torres-guzman_casimir_2006,van_enk_casimir_1995,rodrigues_vacuum-induced_2006}. Another way to understand this phenomena is to realize that photons do not only carry linear momentum but also a nonvanishing angular momentum which can therefore be carried from the vacuum to the objects due to broken rotational symmetry \cite{van_enk_casimir_1995}. More recently there have been several promising proposals for experiments with the goal to measure the Casimir torque between birefrigent materials \cite{munday_torque_2005,iannuzzi_design_2005,chen_measurement_2011}. \\ 
An example of a material which is able to break rotational symmetry and which is of great interest at the moment is provided by topological insulators (TI) \cite{hasan_colloquium_2010}. Topological insulators behave like regular insulators in their bulk but they possess conducting surface states. Originally they were proposed for electronic states but in more recent years it was shown that they also exist in so called photonic topological insulators (PTI) \cite{raghu_analogs_2008,hassani_gangaraj_berry_2017,lu_photonic_nodate,khanikaev_photonic_2012} as for example magnetized plasma \cite{gao_photonic_2016,silveirinha_chern_2015,silveirinha_bulk-edge_2016}. One of the most striking features of TIs is that there exist unidirectional waves on the surfaces of these materials which turn out to be immune to backscattering \cite{davoyan_theory_2013,haldane_possible_2008}. Due to this directionality of the edge states PTIs do not only have a distinguishable axis as i.e. birefrigent materials but their axes possesses also a \textit{direction}, that means an axis that is sensitive to whether an electromagnetic wave is traveling in one or the other direction along the axis. This feature has been of great interest and was used to construct devices like directional wave guides \cite{haldane_possible_2008}, optical isolators or circulators. \\
Now the natural question arises what quantum optical effects emerge when working with PTIs. This has been done by previous authors before: they studied the influence of the presence of a PTI on the entanglement of a two level system \cite{gangaraj_robust_2017}; in Ref.~\cite{silveirinha_fluctuation-induced_2017,gangaraj_spontaneous_2017} the normal and lateral Casimir-Polder force acting on an atom close to a vacuum/PTI interface was analyzed; a huge anisotropic thermal magnetoresistance was obtained in the near field radiative heat transfer between two spherical particles consisiting of a PTI in Ref.~\cite{ekeroth_anisotropic_2017}; but also the Casimir force has already been studied for two infinite half spaces consisting of PTIs, namely InSb, in Ref.~\cite{fuchs_casimir-lifshitz_2017}. All those works showed that there exist interesting new features in quantum optics arising from the interplay of the quantized electromagnetic field with PTIs. However, the unidirectional features of PTIs have not yet been seen manifest in Casimir energies and torques between two macroscopic objects. \\   
One of the main challenges when working with PTIs is that one has to take nonreciprocal material response into account, e.g. arising from the unidriectional surface states, which has been done in Ref.~\cite{buhmann_macroscopic_2012}. Optical effects for nonreciprocal materials have been studied in recent years leading to some stunning results like a persistent directional heat current between three objects at thermal equilibrium \cite{zhu_persistent_2016,silveirinha_topological_2017}. The expression for the Casimir force has also been generalized for the case of nonreciprocal materials and has been applied to PTIs in Ref.~\cite{fuchs_casimir-lifshitz_2017}. 


In this Letter, we want to show how the unidirecitonality and nonreciprocity of PTI plates manifest itself in the Casimir force and torque. Therefore we will show in the following that, in addition to a normal component of the Casimir force, there exists a non-negligible Casimir torque whose magnitude and direction is tuneable by the external magnetic fields. Furthermore, due to the directionality of the topological surface states, we find that this torque is $2\pi$-periodic with respect to the relative angle between the two bias magnetic fields, in sharp distinction from the $\pi$-periodicity occurring for reciprocal bianisotropic media. We also discuss how the tuneability of the Casimir torque can be exploited in nanomechanical schemes to induce rotations. 

\begin{figure}
\centering
\includegraphics[width=0.6\linewidth]{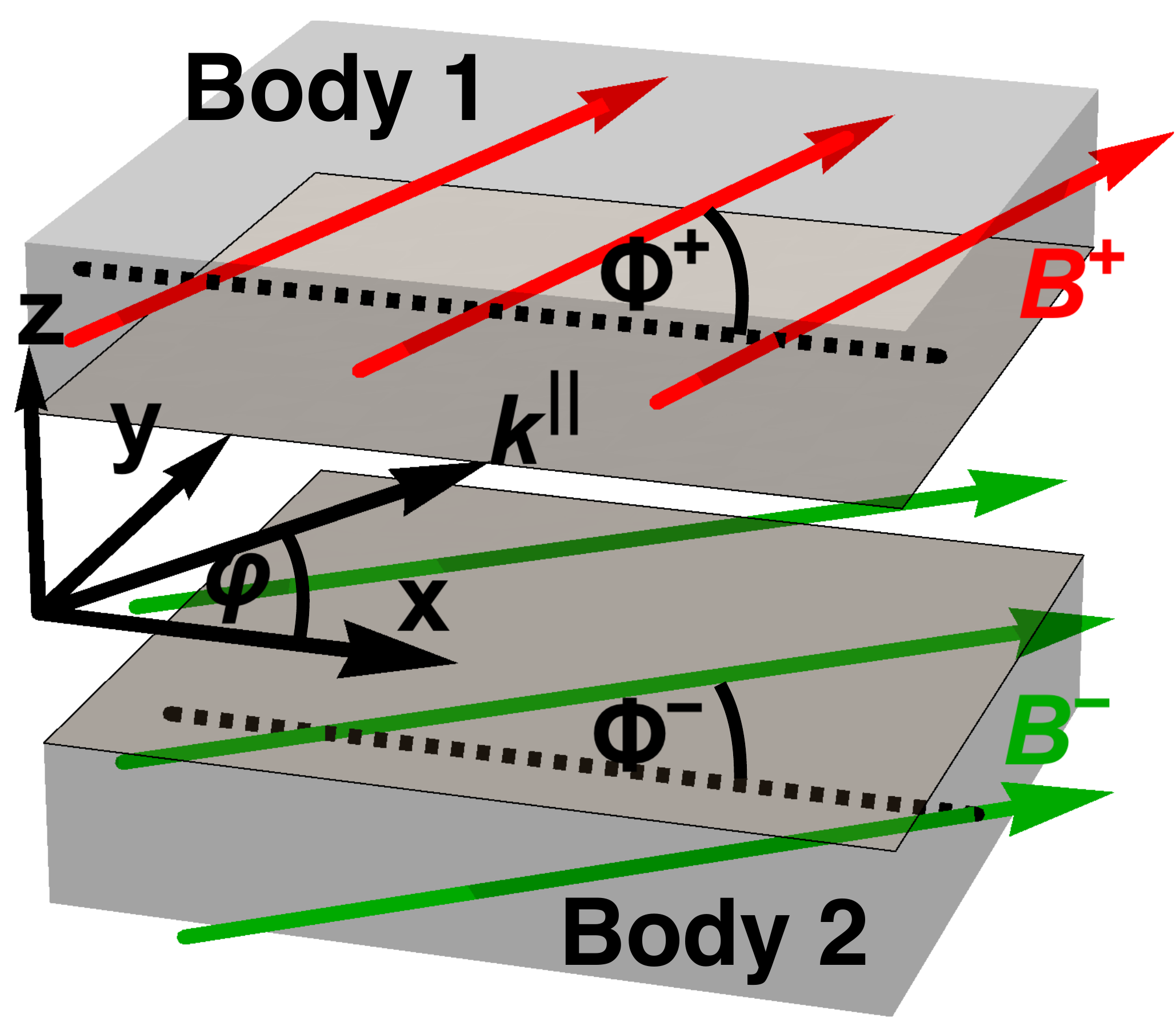}
\caption{We consider a setup as shown here consisting of two semi infinite half spaces filled with a PTI. Additional in each half space there is a applied external magnetic field $\textbf{B}^+$ and $\textbf{B}^-$. The fields are in the $xy-$plane and are parametrised by the angles $\Phi^+$ and $\Phi^-$ which are the angles between $\textbf{B}^+$, $\textbf{B}^-$ and the $x-$axis, respectively. Furthermore, as shown, we define the angle $\varphi$ as the angle between the parallel component $\vec{k}^\parallel$ of the wave vector and the $x-$axis. }
\label{fig:setup}
\end{figure}

We study a setup consisting of two semi infinite half spaces separated by vacuum with separation $L$ and filled with PTIs where a bias magnetic field is applied to each half space (see Fig.~\ref{fig:setup}). The PTI is implemented by a magnetized plasma with an applied bias magnetic field $\vec{B}$ as an example of a gyrotropic material which is described by a permittivity tensor of the form \cite{palik_coupled_1976}
\begin{multline}
\Greektens{$\epsilon$} = \begin{pmatrix} \epsilon_\textrm{xx} & 0 & 0 \\ 0 & \epsilon_\textrm{zz} & \epsilon_\textrm{yz} \\ 0 & - \epsilon_\textrm{yz} & \epsilon_\textrm{zz} \end{pmatrix}, \text{  with} \hspace{4pt} \epsilon_{yz}(\omega) =\frac{ \mi \omega_{\textrm{c}} \omega_{\textrm{p}}^2}{\omega (\omega_{\textrm{c}}^2- \left( \omega + \mi \gamma \right)^2 )}  \\
\epsilon_{zz}(\omega) =	 1 + \displaystyle{\frac{\omega_{\textrm{L}}^2-\omega_{\textrm{T}}^2}{-\mi \Gamma \omega + \omega_{\textrm{T}}^2 - \omega^2}} +\frac{\omega_{\textrm{p}}^2 \left( \omega + \mi \gamma \right)}{\omega ( \omega_{\textrm{c}}^2- \left( \omega + \mi \gamma \right)^2 ) }, \\
\epsilon_{xx}\omega = 1 + \displaystyle{\frac{\omega_{\textrm{L}}^2-\omega_{\textrm{T}}^2}{-\mi \Gamma \omega + \omega_{\textrm{T}}^2 - \omega^2}} -\frac{\omega_{\textrm{p}}^2}{\omega \left( \omega + \mi \gamma \right)}
\label{eq:Epsilon}
\end{multline}
if $\vec{B}$ points in $x$-direction. 
This is easily generalized for arbitrary directions of the magnetic field by simply rotating $\Greektens{$\epsilon$}$. 
The plasma and cyclotron frequencies are given by
$\omega_{\textrm{p}} =	\sqrt{n q_{\textrm{e}}^2/( m^\star \epsilon_0)}$ and 
$\omega_{\textrm{c}} = B q_{\textrm{e}}/m^\star$, respectively, where $q_e$ is the electron charge, $m^\star$ its reduced mass, $n$ is the free electron density and $\gamma$ is the free carrier damping constant. Furthermore $\Gamma$ represents the phonon damping constant and $\omega_L$ and $\omega_T$ are the longitudinal and transverse optic-phonon frequency, respectively. Throughout this paper we will use the following values for the material constants of InSb which have been measured in Ref.~\cite{palik_coupled_1976}: $\omega_{\textrm{L}} = 3.62 \cdot 10^{13}$ \, rad/s, $\omega_{\textrm{T}} = 3.39 \cdot 10^{13}$ \, rad/s, $\Gamma = 5.65 \cdot 10^{11}$ \, rad/s, $\gamma = 3.39 \cdot 10^{12}$ \, rad/s, $n = 1.07 \cdot 10^{17}  \, \text{cm}^{-3}$, $m^\star = 0.022 \cdot m_{\textrm{e}}$ where $m_{\textrm{e}}$ is the electron mass. Furthermore, as depicted in Fig.~\ref{fig:setup}, we define the two applied magnetic fields $\textbf{B}^+$ and $\textbf{B}^-$ laying in the $xy-$ plane by their absolute values $B^+$ and $B^-$ and by the angles between $\textbf{B}^+$ and $\textbf{B}^-$ and the $x-$axis, namely $\Phi^+$ and $\Phi^-$. Although we choose a specific PTI model here, our findings are general and can be applied to other specific PTI realizations. 


We first have to derive a general expression for the Casimir torque of our system. Normally this is done by directly calulating the Casimir energy $E$ to obtain the Casimir torque from it. Nevertheless we are going to first calculate the Casimir force and derive an expression for $E$ from it, since in our setup the permittivity tensor in Eq.~\eqref{eq:Epsilon} breaks the isotropy in the $xy-$plane and therefore there could possibly exist a lateral Casimir Force. This lateral component of the Casimir force would not show up by directly calculating the Casimir energy. To calculate the Casimir force $\vec{F}$ acting on body one we cannot use the final result of Ref.~\cite{fuchs_casimir-lifshitz_2017} (Eq.~(31) in Ref.~\cite{fuchs_casimir-lifshitz_2017}), due to the broken isotropy in the $xy-$plane. Nevertheless we can use the intermediate result Eq.~(18) found in the same reference: 
\begin{multline}
\vec{F} = -\frac{\hbar}{2 \pi} \int\limits^{\infty}_0 \dif \xi \int \limits_{\partial V} \dif \vec{A} \\
\cdot \left\{ \frac{2\xi^2}{c^2} \mathcal{S} \left[ \tens{G}^{(1)} \left( \vec{r}, \vec{r}', \mi \xi \right) \right] + 2 \overrightarrow{\nabla} \times \mathcal{S} \left[ \tens{G}^{(1)} \left( \vec{r}, \vec{r}', \mi \xi \right)\right] \times \overleftarrow{\nabla}'  \right.  \\
\left. -\textrm{Tr} \left[ \frac{\xi^2}{c^2} \tens{G}^{(1)} \left( \vec{r}, \vec{r}', \mi \xi \right) + \overrightarrow{\nabla} \times \tens{G}^{(1)} \left( \vec{r}, \vec{r}', \mi \xi \right) \times \overleftarrow{\nabla}' \right] \tens{1} \right\}_{\vec{r}' \rightarrow \vec{r}}.
\label{eq:Casimir Force 1}
\end{multline}
Here we introduced the symmetrization $\mathcal{S}$ of a tensor defined by $\mathcal{S} \left[ \tens{G} \left( \vec{r}, \vec{r}', \omega \right) \right] = (1/2) \left[ \tens{G} \left( \vec{r}, \vec{r}', \omega \right) + \tens{G}^{\textrm{T}} \left( \vec{r}', \vec{r}, \omega \right) \right]$. Furthermore $\xi = - \mi  \omega$ where $\omega$ is the frequency of the electromagnetic wave, $\partial V$ is any infinite planar surface in the vacuum gap between the two planar bodies and $\dif \vec{A}$ its surface element, $\vec{r}$ and $\vec{r}^\prime$ are arbitrary points on the surface of body one. Most importantly $\tens{G}^{(1)} \left( \vec{r}, \vec{r}', \mi \xi \right)$ is the scattering Greens tensor \cite{buhmann_dispersion_2012} of our setup. $\tens{G}^{(1)} \left( \vec{r}, \vec{r}', \mi \xi \right)$ can be expressed using the reflection coefficients of the vacuum/PTI interfaces calculated in the supplementary material ?? and the different components of the wave vector $\vec{k} = (\vec{k}^\parallel,k_z=\mi \kappa)^\textrm{T}$ satisfying $\vec{k}^2 = -\xi^2/c^2$ where $c$ is the speed of light in vacuum. The full expression of $\tens{G}^{(1)} \left( \vec{r}, \vec{r}', \mi \xi \right)$ and its derivation can also be found in the supplementary material ??. Note that in Ref.~\cite{silveirinha_fluctuation-induced_2017} an expression for $\tens{G}^{(1)} \left( \vec{r}, \vec{r}', \mi \xi \right)$ in case of a setup with only one PTI half space has already been calculated.  \\
Using these results we find first of all, as expected, that there is no lateral force in the ground state of the system and thus the $x$ and $y$ components of $\vec{F}$ are zero. Furthermore in the supplementary material ?? we find a general expression for the normal component of $\vec{F}$ per unit area $A$ which we define as $f \equiv F_z/A$, where $F_z$ is the $z-$component of $\vec{F}$. This result is not shown here since we are only interested in the near field behavior of our system. Thus we want to analyze $f$ further under the assumption $\xi/c \ll k^\parallel$ which is often referred to as the nonretarded limit which becomes valid for small separations $L$, compare Ref.~\cite{buhmann_dispersion_2012}. Under this assumption the reflection coefficients simplify significantly to $r_{\textrm{s}, \textrm{s}}^\pm  \simeq r_{\textrm{s}, \textrm{p}}^\pm  \simeq r_{\textrm{s}, \textrm{p}}^\pm  \simeq 0$ and 
\begin{multline}
\label{rp}
r_{\textrm{pp}}^\pm(\omega)   \simeq \frac{-2  + D \mp 2 \mi \epsilon_{yz} \sin(\varphi - \Phi^\pm )}{2  + D \mp 2 \mi \epsilon_{yz} \sin(\varphi - \Phi^\pm)}, \quad \text{where}\\
D = \sqrt{2\epsilon_{zz}(\epsilon_{\textrm{zz}}+\epsilon_{\textrm{xx}} + (\epsilon_{\textrm{xx}} - \epsilon_{\textrm{zz}})\cos\left[2(\varphi - \Phi^\pm)\right])}.
\end{multline}
Here $\varphi$ is defined by $\vec{k}^\parallel =k^\parallel \left( \sin(\varphi), \cos(\varphi) \right)^\textrm{T}$ with $k^\parallel \equiv |\vec{k^\parallel}|$ (compare Fig.~\ref{fig:setup}) and the "$^\pm$" indicates the reflection at the lower and upper half space, respectively. 
Note that this reflection coefficient is not real even when evaluated at imaginary frequencies $\omega=\mi \xi$ due to the terms proportional to $\epsilon_{\textrm{yz}}$. But since this term also flips sign under $\vec{k}^\parallel \to - \vec{k}^\parallel$ the Schwarz reflection principle $\tens{G}^{(1)} \left( \vec{r}, \vec{r}', \mi \xi \right) = \tens{G}^{(1)\star} \left( \vec{r}, \vec{r}', \mi \xi \right) $ is obeyed which according to \cite{buhmann_impact_2016} implies $r_{\textrm{pp}}^\star(\vec{k}^\parallel, \mi \xi) = r_{\textrm{pp}}(-\vec{k}^\parallel, \mi \xi)$. Thus it is ensured that the Greens tensor and therefore the Casimir force is real. \\
Finally using the previous result of the reflection coefficients in the nonretarded limit $f$ simplifies to
\begin{multline}
f = - \frac{\hbar}{16 \pi^3 L^3} \int\limits_0^\infty \dif \xi \int\limits_0^{2\pi} \dif \varphi \int\limits_{0}^\infty \dif \kappa \hspace{2pt} \kappa^2  \frac{r_\textrm{pp}^+(\mi \xi)r_\textrm{pp}^-(\mi \xi) \me^{-2\kappa L} }{1-r_\textrm{pp}^+(\mi \xi)r_\textrm{pp}^-(\mi \xi) \me^{-2\kappa L}}  \\
= - \frac{\hbar}{16 \pi^3 L^3} \int\limits_0^\infty \dif \xi \int\limits_0^{2\pi} \dif \varphi \hspace{4pt}  \textrm{Li}_{3}\left[r_\textrm{pp}^+(\mi \xi)r_\textrm{pp}^-(\mi \xi) \right]
\label{eq:force final}
\end{multline}
where $\textrm{Li}_{3}$ is the polylogarithm of order three. As in the reciprocal case \cite{buhmann_dispersion_2012} and for a magnetized plasma with a bias magnetic field perpendicular to the interface \cite{fuchs_casimir-lifshitz_2017} we find a simple $f \propto 1/L^3$ behavior in the nonretarded limit. Therefore we can easily calculate the Casimir energy $E$ per unit area in the nonretarded limit from Eq.~\eqref{eq:force final} by integrating $f$ with respect to $L$ under the boundary condition $E \to 0$ if $L \to \infty$ and eventually find $E= L\,f/2$.
From this result we can now calculate the Casimir torque $T = -\partial E /(\partial \Delta \Phi) $ where $\Delta \Phi = \Phi^- - \Phi^+$. \\

\begin{figure}
\centering
  \subfigure{\includegraphics[width=0.47\linewidth]{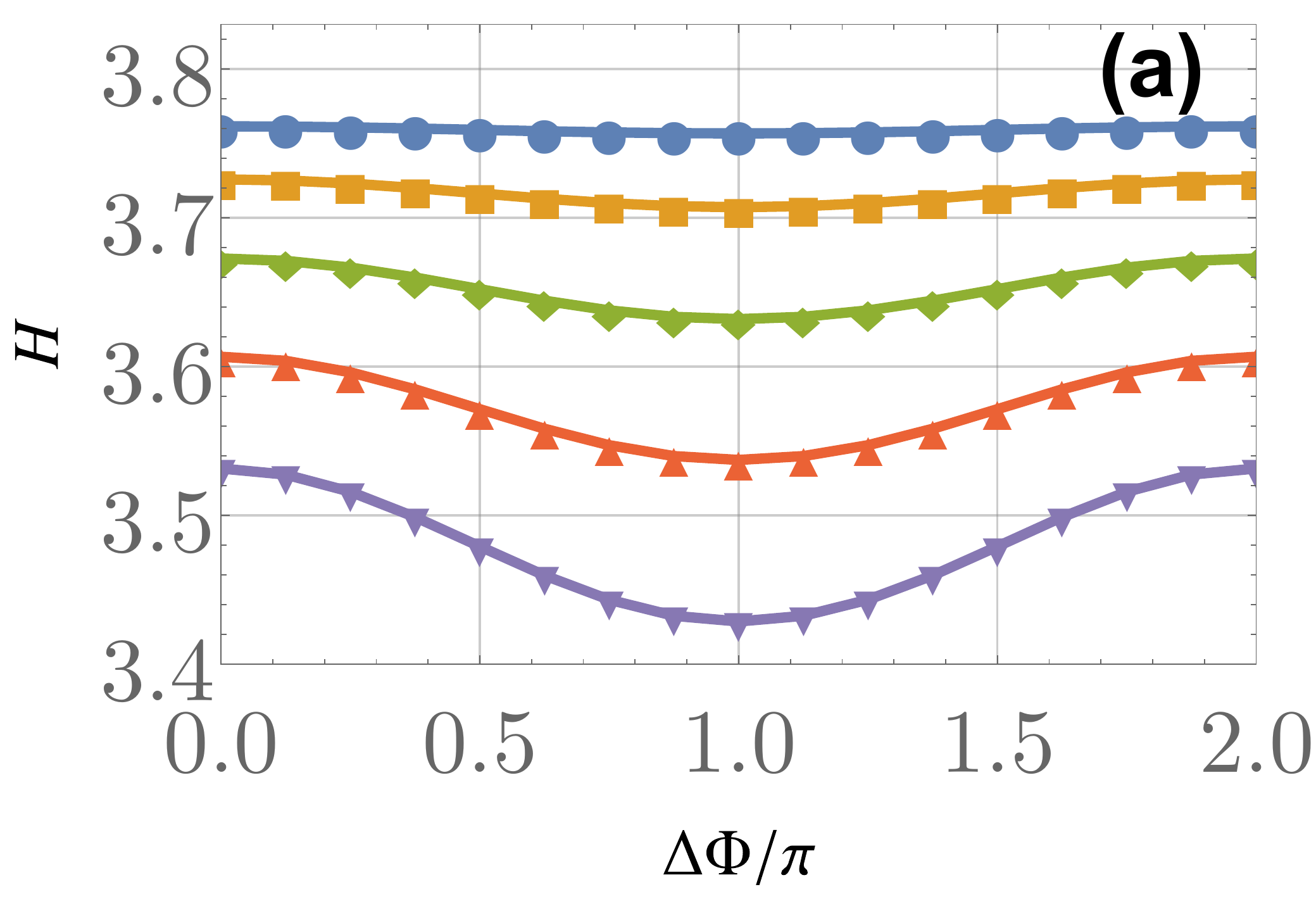}}
   \subfigure{\includegraphics[width=0.5\linewidth]{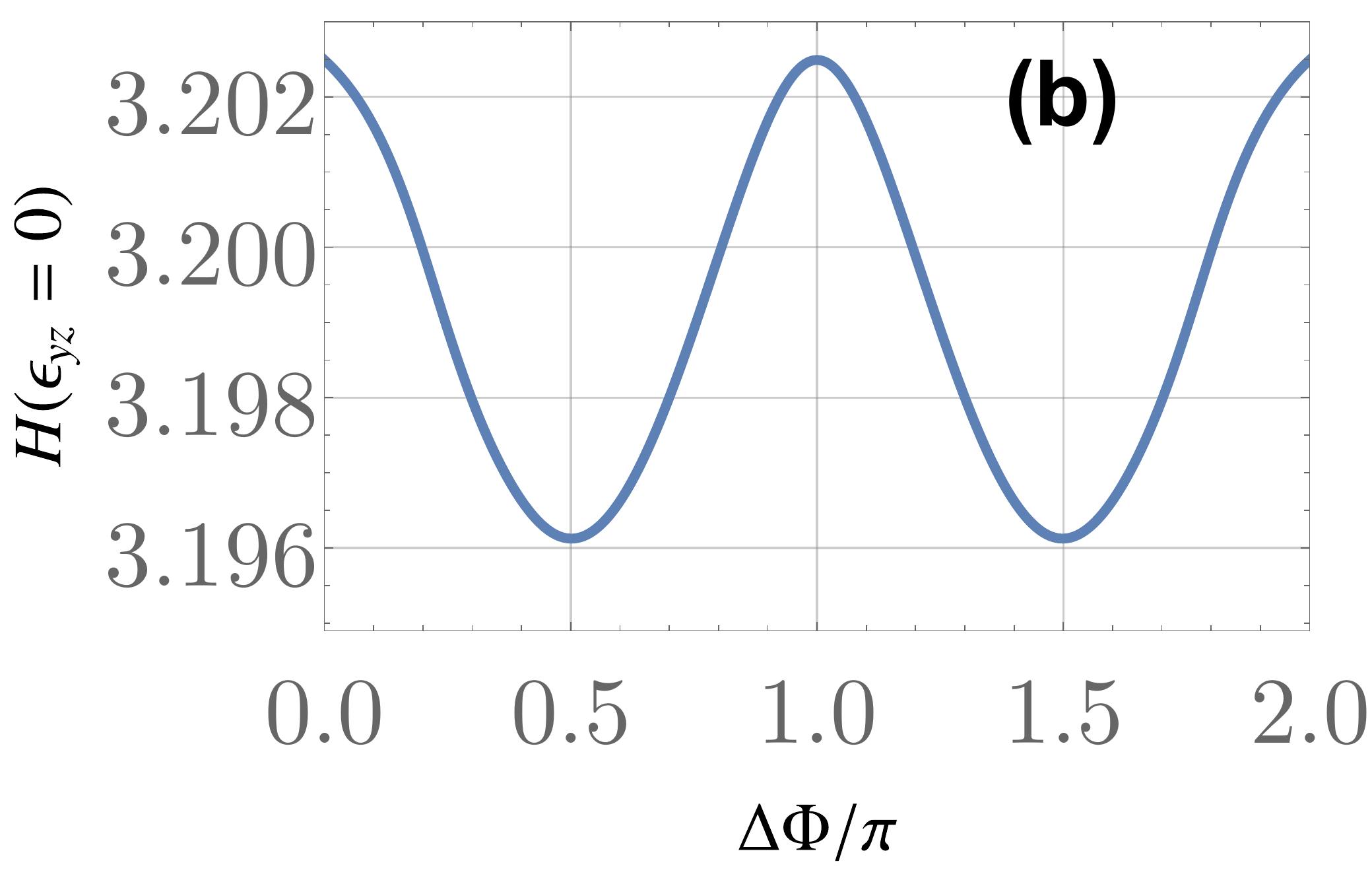}} 
   \subfigure{\includegraphics[width=0.48\linewidth]{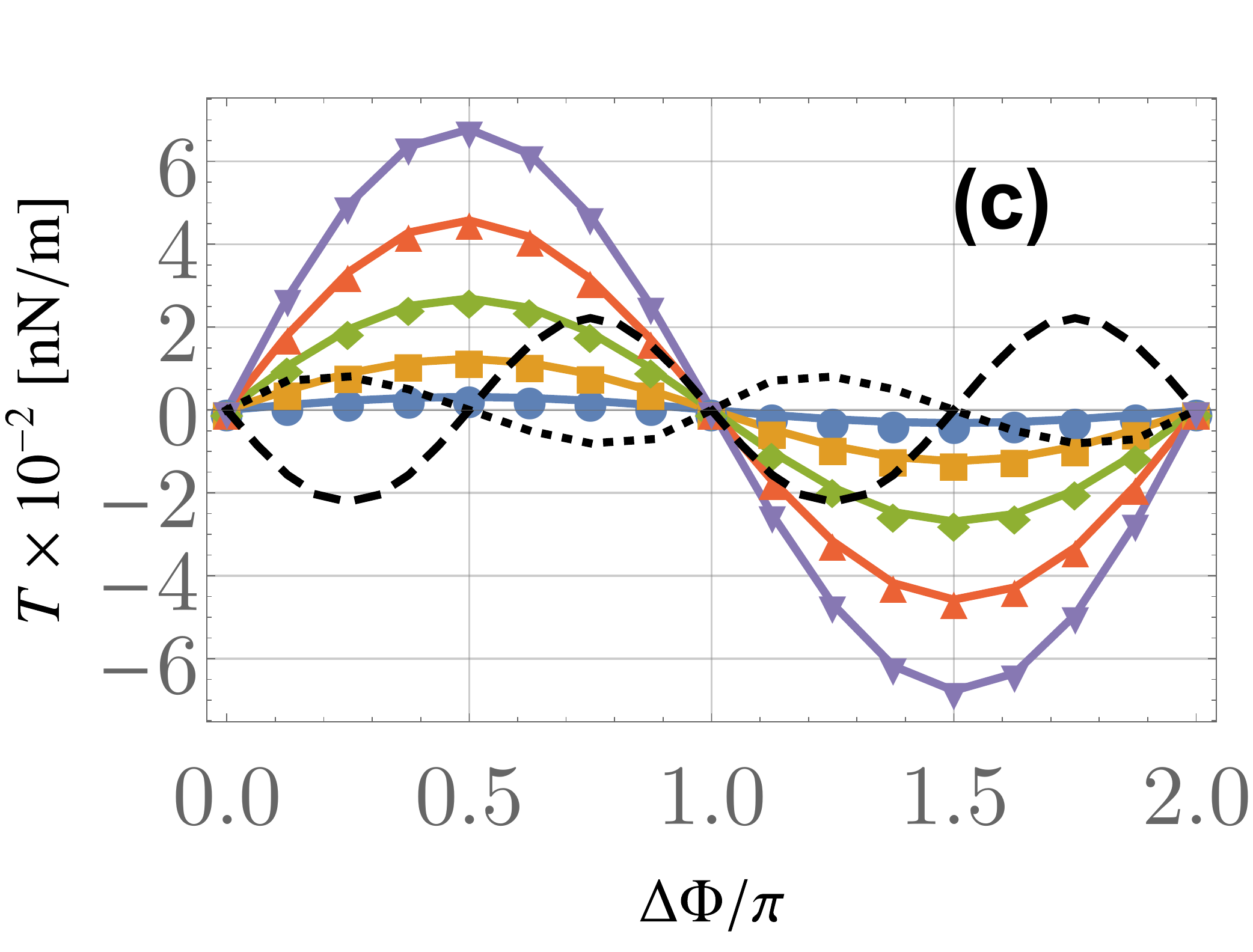}}
   \subfigure{\includegraphics[width=0.48\linewidth]{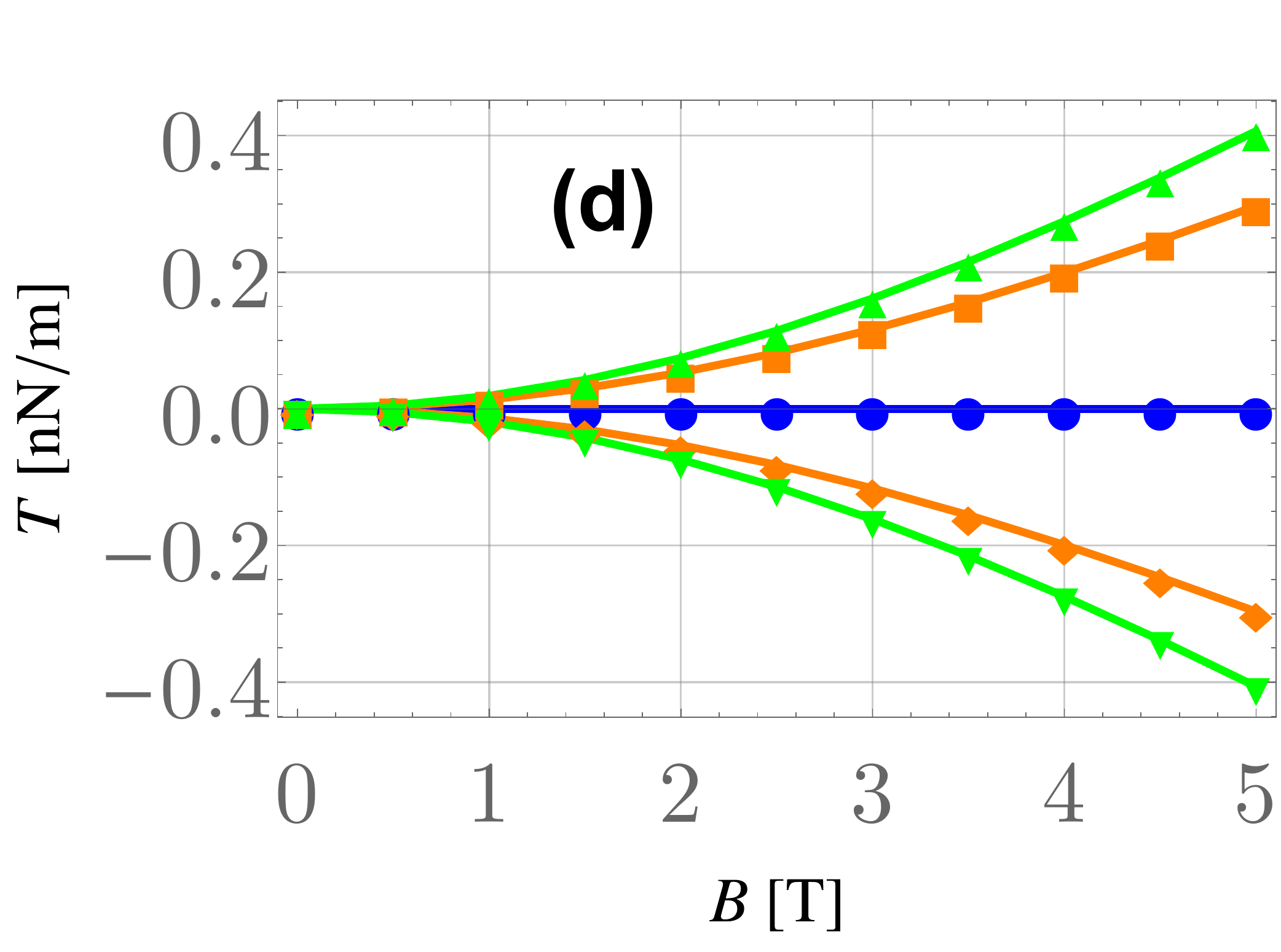}}  \vspace{-10pt}
\caption{Dimensionless Casimir energy, torque and force. In (a) we plot the dimensionless Hamaker constant $H$ as a function of $\Delta \Phi$ for different values of $B= 1,2,3,4,5$\,T (dots, squares, diamonds, triangles, upside down triangles). In (b) we used again the $\epsilon$-tensor in Eq.~\eqref{eq:Epsilon} with the only exception that we set $\epsilon_\textrm{yz} =0$ to obtain a description of a normal anisotropic medium ($B=5$\,T). The Casimir torque $T(\Delta \Phi)$ is plotted in (c) for different magnetic fields where the same shapes correspond to the same value of $B$ as in (a). Additionally we plotted the Casimir torque for the material model of plot (b) (dotted line) and for a setup where the two PTI half spaces are replaced by one barium titanate and one quartz half space (dashed line). Finally we plotted the Casimir torque as a function of $B$ in (d) for different values of $\Delta \Phi= 0,\pi/4,\pi/2,5\pi/4,3\pi/2 $ (dots, squares, triangles, diamonds, upside down triangles). In all plots we used $L=100$\,nm.}
\label{fig:Energy and Force}
\end{figure}

Next, we want to analyze the previous results for the Casimir force, energy and torque. To this end, in Fig.~\ref{fig:Energy and Force}, we display the dimensionless Hamaker constant defined by $H \equiv \int_0^\infty (\dif \xi/\omega_p) \int_0^{2\pi} \dif \varphi \hspace{4pt}  \textrm{Li}_{3}\left[r_\textrm{pp}^+r_\textrm{pp}^- \right]$ at a fixed gap distance $L = 100$\,nm. Note that one can easily retrieve $f$, $E$ and $T$ from $H$ via $f = -\omega_p\hbar H/16 \pi^3 L^3$, $E=-\omega_p\hbar H/32 \pi^3 L^2$ and $T=(\omega_p\hbar /32 \pi^3 L^2) \partial H /(\partial \Delta \Phi)$. Before discussing the qualitative features of these results let us mention a few things about the magnitude of the torque. As is depicted in Fig.~\ref{fig:Energy and Force} (c) the Casimir torque at zero temperature for two semi infinite PTI half spaces reaches the same order of magnitude as the one for quartz or calcite half spaces kept parallel to a barium titanate half space. Those examples for birefringent plates have been studied in Ref.~\cite{munday_torque_2005} and we have used the same model for the permittivity including the same values for the constants measured in Ref.~\cite{bergstrom_hamaker_1997} to reproduce these results. More concretely this means that the torque for the PTI setup reaches a maximal torque of about $67$\,pN/m with $B=5$\,T whereas the quartz (calcite) - barium titanate setup reaches $22$\,pN/m ($317$\,pN/m). Nevertheless, we remark that the Casimir torque between two corrugated metals is three orders of magnitude larger \cite{rodrigues_vacuum-induced_2006}. But note that these torques are all periodic under a rotation of $\pi$ of one of the plates around its normal component since their distinguished axes are not directional.   \\
Thus, now we want to study the qualitative characteristics of the torque for our setup and we see in Fig.~\ref{fig:Energy and Force} (a) that $H(\Delta \Phi)$ is $2\pi$ periodic with a maximum (minimum) when the two magnetic fields $B^\pm$ point in the same (opposite) direction. 
 This result is therefore qualitatively different from the ones observed when dealing with birefringent half spaces with one in-plane optical anisotropy \cite{munday_torque_2005} (compare the dashed line in Fig.~\ref{fig:Energy and Force} (c)) or corrugated metals \cite{rodrigues_vacuum-induced_2006}. Heuristically, this new periodicity can be explained by the fact that our material model does not only have a distinguished axis, but this axis also has a direction. 

The angle-dependence of the Casimir energy can be understood in more detail by studying the contributions of different surface-plasmon polaritons (SPPs) which dominate in the nonretarded limit. To this end, we take a closer look at the spectral decomposition of the Casimir energy evaluated at real frequencies $\tilde{H}(\omega) \equiv \int_0^{2\pi} \dif \varphi \hspace{4pt} \textrm{Im} \left[ \textrm{Li}_3 \left[ r_\textrm{pp}^+(\omega)r_\textrm{pp}^-(\omega) \right] \right] $ which allows us to see which surface modes contribute the most to the Casimir energy. The total Casimir energy is simply the integral over this spectral energy density, $H= \int_0^\infty (\dif \omega/\omega_p) \tilde{H}(\omega)$ and therefore $E, f  \propto \int_{0}^\infty \dif \omega \hspace{3pt} \tilde{H}(\omega) $, as can be seen from contour-integral techniques and using the Schwartz reflection principle as well as the fact that $\lim_{\omega \to \infty} \omega^2 \tens{G}^{(1)}  = 0$.

\begin{figure}
\centering
\subfigure{\includegraphics[width=0.48\linewidth]{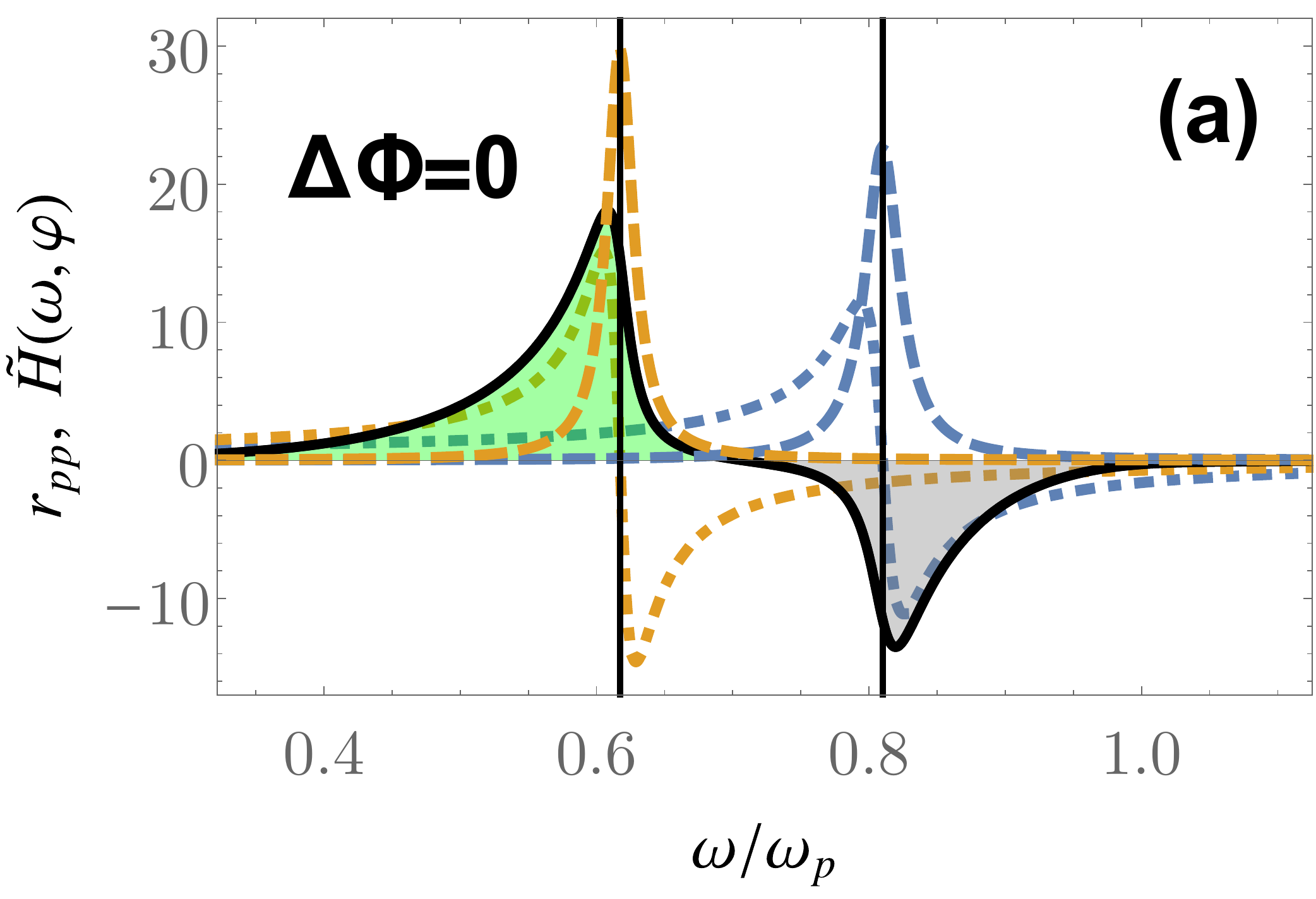}}\hspace{5pt}	   
\subfigure{\includegraphics[width=0.48\linewidth]{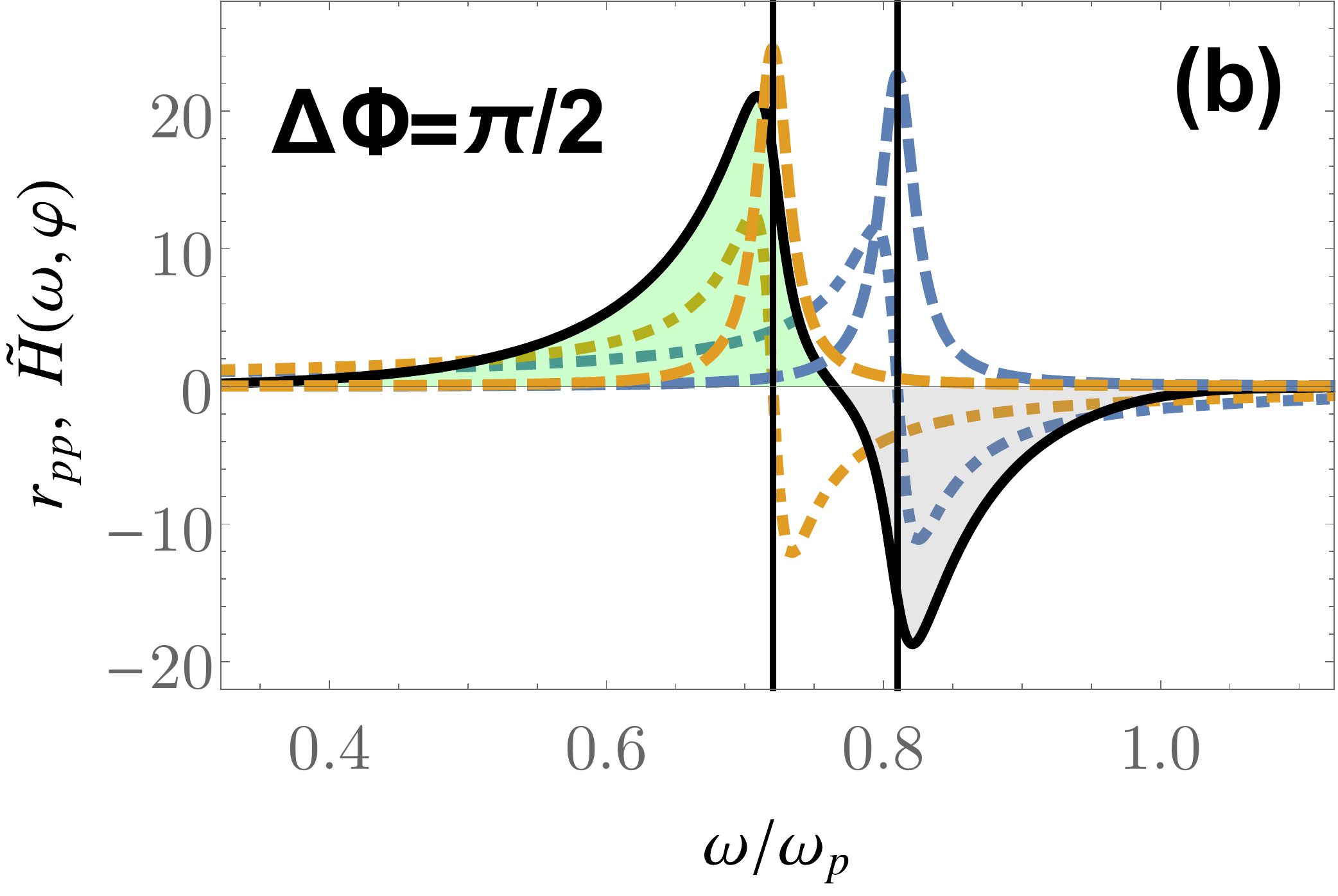}}\hspace{5pt}	
\subfigure{\includegraphics[width=0.48\linewidth]{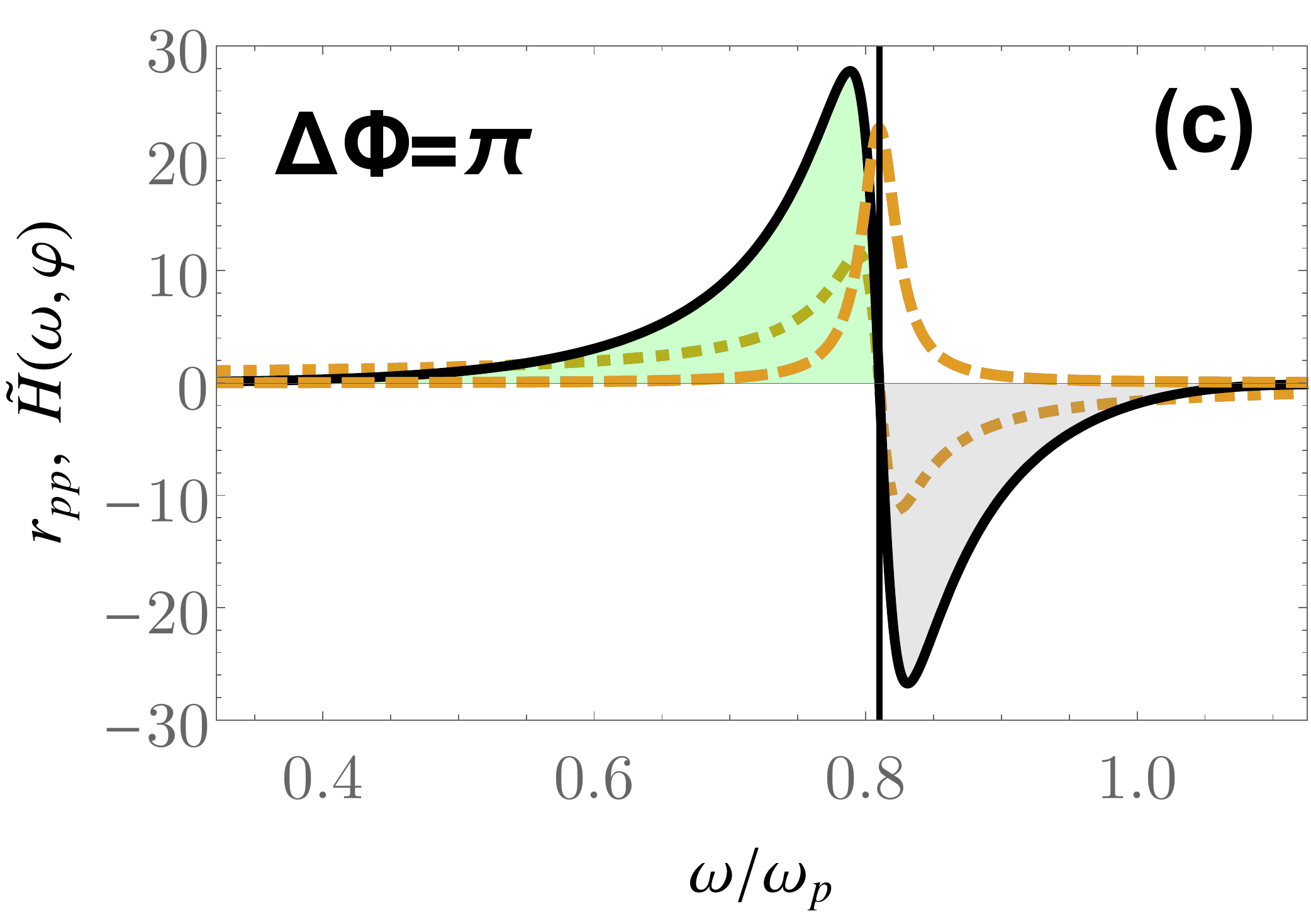}} \hspace{5pt}	    
\subfigure{\includegraphics[width=0.48\linewidth]{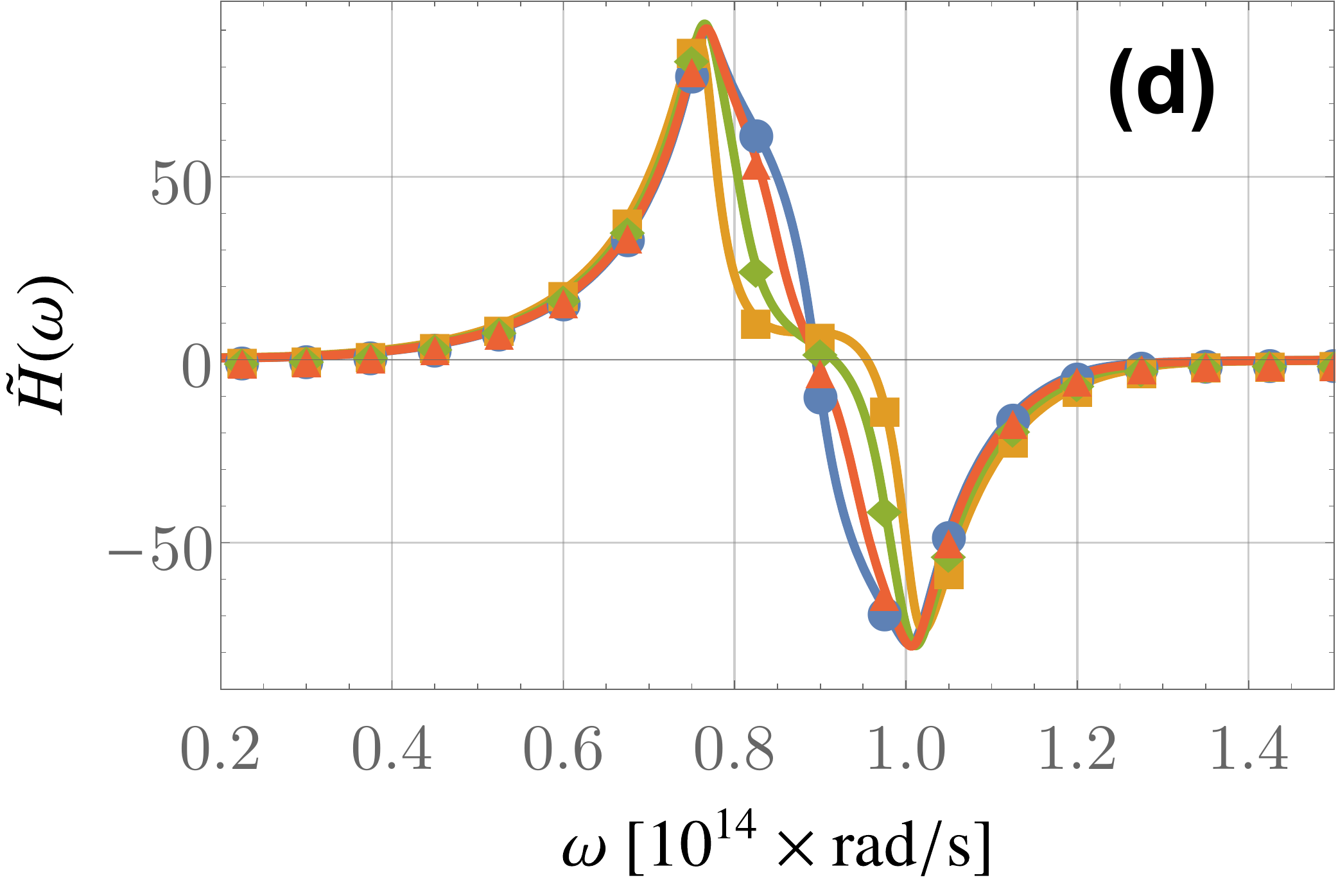}}                      \vspace{-12pt}
		\caption{Contributions of surface-plasmon polaritons (SPPs) to the Casimir energy. (a) Total spectral energy density $\tilde{H}(\omega)$ with $\Delta \Phi = 0,\pi/4,\pi/2,\pi$ (dots, triangles, diamonds, squares). (b)-(c) Single-plate SPPs $\text{Re}\left[r_\textrm{pp}^\pm(\omega)\right]$ (dotted lines) and $\text{Im}\left[r_\textrm{pp}^\pm(\omega)\right]$ (dashed lines) and angle-resolved spectral energy density $\tilde{H}(\omega, \varphi)$ (solid lines) with $ \varphi - \Phi^+ = \pi/2$ and $\Delta \Phi = 0, \pi/2, \pi$ ((b),(c),(d)). $\text{Re}\left[r_\textrm{pp}^+(\omega)\right]$ and $\text{Im}\left[r_\textrm{pp}^+(\omega)\right]$ corresponds to the curves with the resonance at higher frequencies whereas $\text{Re}\left[r_\textrm{pp}^-(\omega)\right]$ and $\text{Im}\left[r_\textrm{pp}^-(\omega)\right]$ correspond to the once with a lower resonance frequency ($\Omega^+>\Omega^-$ in this case); the two single-plate SPP frequencies $\Omega^\pm(\varphi-\Phi^\pm)$ as given by Eq.~\eqref{eq:resonance frequency} are indicated as vertical lines. For all plots we have used $B=3$\,T.}
\label{fig:modes}
\end{figure}

Central ingredient to the spectral Casimir energy density are the SPPs of the individual plates which are resonances of the respective reflection coefficients $r_\textrm{pp}^+(\omega)$ and
$r_\textrm{pp}^-(\omega)$. The frequencies of the SPPs are easily found by setting the denominators in the reflection coefficients~(\ref{rp}) to zero which upon using Eq.~(\ref{eq:Epsilon}) and neglecting the photon contribution leads to the dispersion relations 
\begin{multline}  
\Omega^{\pm}(\varphi-\Phi^\pm) = \frac{1}{4} \left( \sqrt{
    6 \omega_{c}^2 + 8 \omega_p^2 + 2\omega_{c}^2 \cos\left[2( \varphi - \Phi^{\pm})\right]} \right. \\
   \left.  \pm  2 \omega_{c} \sin\left[\varphi - \Phi^{\pm}\right]\right)
\label{eq:resonance frequency}
\end{multline}
in the lossless limit, as also found in  Ref.~\cite{silveirinha_fluctuation-induced_2017}. The SPP frequencies hence depend on the angle between the wave vector and the respective magnetic field, $\varphi-\Phi^\pm$. Thus, for each plate, there is a whole manifold of SPPs at different frequencies for different directions, in contrast to the case of simple metal plates \cite{henkel_coupled_2004}. The single-plate SPPs are illustrated by the dotted and dashed lines in Figs.~\ref{fig:modes}(a)--(c) for selected combinations of magnetic-field and wave vector directions. In particular, we see that the SPP frequencies of the two plates coincide in the special case $\Delta \Phi = \pi$,  Fig.~\ref{fig:modes}(c).

When the two plates are brought in close proximity, as is the case in the nonretarded limit considered, the single plate SPPs combine to form symmetric and antisymmetric coupled SPPs \cite{henkel_coupled_2004}. Mathematically, this can be seen from the Casimir spectral energy density using  Li$_3\left[z \right] \approx \zeta(3)\, z$ for $z\ll 1$ where $\zeta$ is the Riemann zeta function: 
\begin{multline}
\tilde{H}(\omega)  \approx \zeta(3) \left( \textrm{Im}\left[r_\textrm{pp}^+(\omega) \right] \textrm{Re}\left[r_\textrm{pp}^-(\omega) \right] \right.\\ \left. + \textrm{Im}\left[r_\textrm{pp}^-(\omega) \right] \textrm{Re}\left[r_\textrm{pp}^+(\omega) \right] \right)
\label{eq:H approximation}
\end{multline}
We can clearly see how $H(\omega,\varphi)$ is built up from the products $\textrm{Im}\left[r_\textrm{pp}^\pm(\omega) \right] \textrm{Re}\left[r_\textrm{pp}^\mp(\omega) \right]$ in Figs.~\ref{fig:modes}(a)--(c).
The symmetric and antisymmetric coupled SPPs have parallel and antiparallel electric-field vectors on the two surfaces, their frequencies are given by the minimum or maximum of the the two resonances of $\textrm{Im}\left[r_\textrm{pp}^+(\omega) \right] \textrm{Re}\left[r_\textrm{pp}^-(\omega) \right]$ and $\textrm{Im}\left[r_\textrm{pp}^-(\omega) \right] \textrm{Re}\left[r_\textrm{pp}^+(\omega) \right]$, respectively. As illustrated by the solid lines in Figs.~\ref{fig:modes}(a)--(c), the symmetric coupled SPPs give the dominant positive contribution to the Casimir energy (left peak) while the antisymmetric coupled SPPs give a smaller negative contribution (right dip). As further seen in the figures, the difference between the positive and negative contributions is quite pronounced for $\Delta \Phi = 0$ where the SPPs of the two plates and subsequently also the symmetric and antisymmetric SPPs have their largest possible splitting in frequency space, Fig.~\ref{fig:modes}(a), leading to a large net Casimir energy. The splitting is reduced for larger angles, Fig.~\ref{fig:modes}(b) until eventually the single-plate SPPs coincide for  
$\Delta \Phi = \pi$ and the two coupled SPPs become very close in frequency and similar in magnitude. For this case, we have a smaller Casimir energy.    

Our observations remain valid for general combinations of magnetic-field and wave vector directions and thus also when integrating over all wave-vector directions $\varphi \in \left[0,2\pi\right]$ to obtain the total spectral energy density $\tilde{H}(\omega)$. As seen from Fig.~\ref{fig:modes}(d), this energy density has quite a complex profile as it is the sum over contributions from many SPPs with different resonant frequencies. Nevertheless, we again find that positive and negative contributions are most different in magnitude for $\Delta \Phi = 0$ (dots), leading to a large total Casimir energy. As the angle difference increases towards $\Delta \Phi = \pi$ (squares), the positive and negative contributions become more similar in magnitude and the total Casimir energy decreases.

\begin{figure}
\centering
\subfigure{\includegraphics[width=0.4\linewidth]{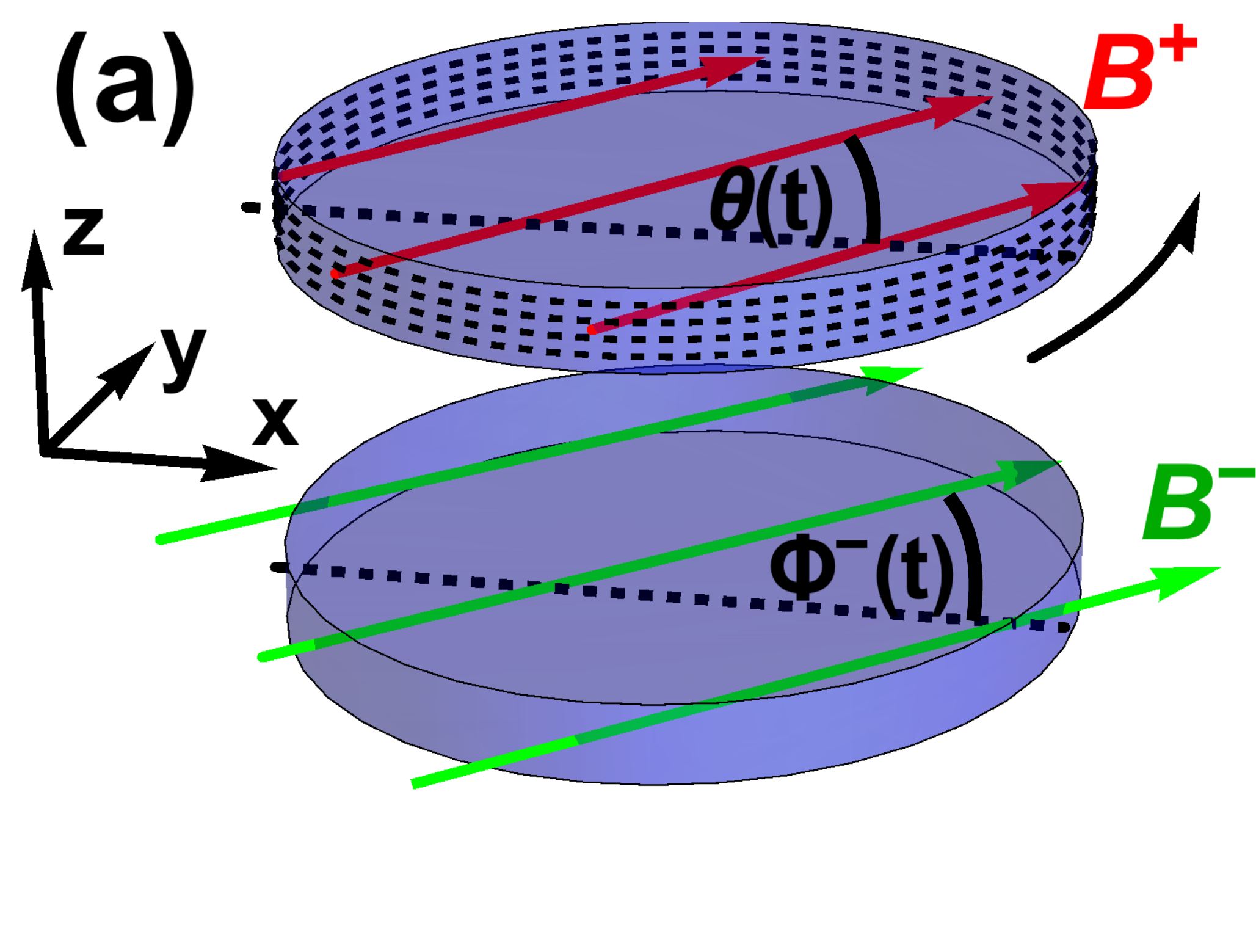}}
 \subfigure{\includegraphics[width=0.5\linewidth]{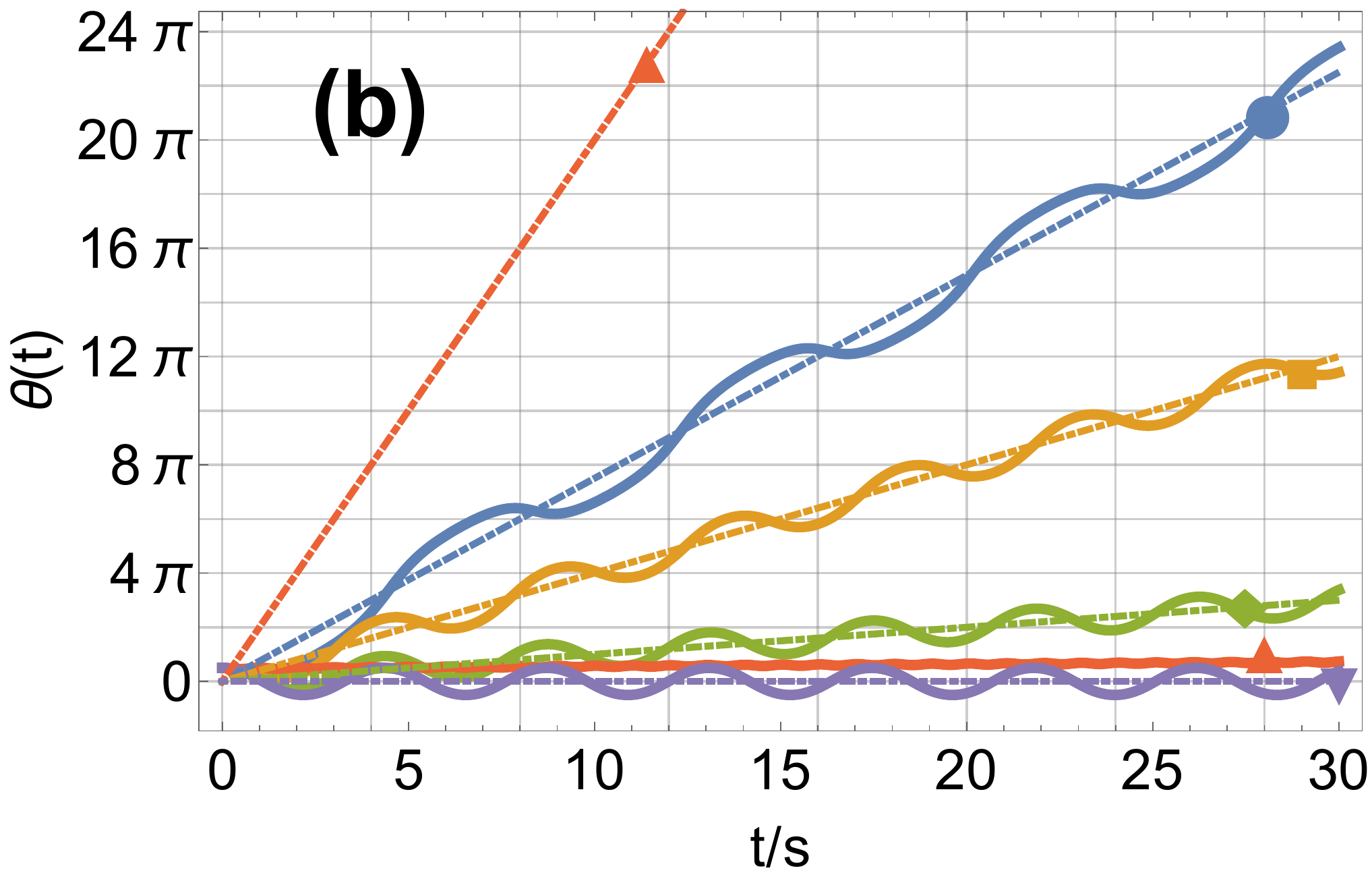}} 
\caption{In (a) one can find the setup under consideration. As shown, there are two disks with radius $r= 20\,\mu$m and thickness $d=20\,\mu$m consisting of InSb and held at a distance of $L=100$\,nm from each other. Furthermore $\vec{B}^+$ is a static magnetic field whose angle with the $x$-axis $\Phi^+$ only changes if the whole disk rotates. The rotation of the upper disk is described by the angle $\theta(t) \equiv \Phi^+(t)$. In the lower disk there is a magnetic field $\vec{B}^-$ whose direction described by the angle $\Phi^-(t)$ may change over time although the plate is fixed. In (b) we show the numerical solution to Eq.\eqref{eq:eom} with $B^+=B^-=5$\,T for the cases $\Phi^-(t) = 0.75\pi t, 0.4 \pi t, 0.1\pi t, 2\pi t, 0$ (dot, square, diamond, triangle, upside down triangle) with solid lines. Additionally we plot $\Phi^-(t) = 0.75\pi t, 0.4 \pi t, 0.1\pi t, 2\pi t, 0$ (indicated also by dot, square, diamond, triangle, upside down triangle) describing the rotation of  $\vec{B}^-$ with dotted-dashed lines, to see how the upper disk follows $\vec{B}^-$ if the angular velocity of $\Phi^-(t)$ is smaller or equal to $0.75$\,rad/s.}
\label{fig:eom}
\end{figure}
The advantage of having an in situ tuneability of the torque can be exploited for nanomechenical schemes. Therefore we are going to show how one can set a disk consisting of InSb into rotation via a time dependent bias magnetic field. \\
For instance, we consider a setup as depicted in Fig.~\ref{fig:eom} (a), where two disks with the same size as considered in Ref.~\cite{iannuzzi_design_2005} namely with radius $r= 20\,\mu$m and thickness $d=20\,\mu$m consisting of InSb which has a mass density of $\rho_{\text{InSb}}=5.59\,\text{g}/\text{cm}^3$ \cite{bateman_elastic_1959} are held at a fixed distance from another. Additionally, in the upper disk there is a constant magnetic field $\vec{B}^+$ applied which is attached to the disk e.g. via magnetic coating \cite{grushin_effect_2011}. This disk is free to rotate around the $z$-axis and therefore the angle between $\vec{B}^+$ and the $x$-axis, namely $\Phi^+ \equiv \theta(t)$ may change over time due to the rotation of the disk. The lower disk is fixed and thus not free to rotate. Furthermore in the lower disk there exists a magnetic field  which is not attached to it and whose angle with the $x$-axis $\Phi^-(t)$ is tuneable. Therefore the relative angle between the two magnetic fields is given by $\Delta\Phi(t) = \Phi^-(t) - \theta^+(t)$.
The dependence of the torque $T$ on the relative angle between the applied magnetic fields is very well approximated by $T(\Delta\Phi) \cong T_0 \sin(\Delta\Phi)$ (compare Fig.~\ref{fig:Energy and Force} (c)) where $T_0 = 67\,$pN/m if $B^+=B^-=5$\,T. Thus the equation of motion for the rotation of the upper disk neglecting finite size effects and friction is given by
\begin{equation}\label{eq:eom}
\frac{\dif^2 \theta}{\dif t^2} = \frac{2T_0}{l r^2 \rho_\text{InSb}}  \sin\left[ \Phi^-(t)- \theta^+(t) \right].
\end{equation}   \\
Here $r$ and $l$ are the radius of the upper disk and its thickness, respectively. 
The result of the numerical solution of Eq.~\eqref{eq:eom} can be found in Fig.~\ref{fig:eom} (b). As one can see if we let the tunable magnetic field $\vec{B}^-$ rotate with an angular velocity of up to $0.75$\,rad/s the upper plate will follow the direction of $\vec{B}^-$ and start to rotate with the same angular velocity of almost one full $2\pi$ rotation per second. If $\vec{B}^-$ rotates faster than $0.75$\,rad/s the upper disk can not follow the direction of $\vec{B}^-$ and therefore it does barely rotate (triangle).
 

To summarize, we have found a Casimir Torque between topological-insulator plates whose direction and magnitude is easily tuneable by the external bias magnetic field. Furthermore, the torque is of the same order of magnitude as the ones between quartz or calcium and barium titinate half spaces. We have further shown that in the nonretarded limit this torque is dominated by SPPs which are directional and therefore it is only symmetric under a rotation of $2\pi$ of one of the plates around its normal component. This unique periodicity in contradistinction from the typical $\pi$-periodicity for ordinary birefringent media is a clear signature of nonreciprocity. 
We have shown how the tuneability of the torque between two InSb disks can be exploited to set one of the disks into rotation which offers new possibilities for measurements and nanomechanical applications of the Casimir torque on small objects. 

We acknowledge helpful discussions with Francesco Intravaia and Alexey Belyanin. This work was supported by the German Research Foundation (DFG, Grants BU 1803/3-1 and GRK 2079/1). S.Y.B is grateful for support by the Freiburg Institute of Advanced Studies.


%



\end{document}